\documentstyle[psfig,12pt,aaspp4]{article}
\def\kms{km~s$^{-1}$}
\def\hal{H$\alpha$}
\def\be{\begin{equation}}
\def\ee{\end{equation}}
\def\about{$\sim$}

\def\h{$h^{-1}$}
\def\HI{\ion{H}{1}}
\def\HII{\ion{H}{2}}
\def\NII{[\ion{N}{2}]}
\def\SII{[\ion{S}{2}]}
\def\Wobs{$W_{\rm obs}$}
\def\Ropt{$R_{\rm opt}$}

\begin{document}
\hskip 3.5in{\hskip 10pt \date{13 November 1999}}
\title{Signatures of Interstellar-Intracluster Medium
Interactions: Spiral Galaxy Rotation Curves in Abell 2029}
 
\author {DANIEL A. DALE}\affil{IPAC, California Institute of Technology
100-22, Pasadena, CA 91125}

\author {JUAN M. USON}\affil{National Radio Astronomy Observatory,
520 Edgemont Road, Charlottesville, VA 22903}

\begin{abstract}
We investigate the rich cluster Abell~2029 ($z \sim 0.08$) using 
optical imaging and long-slit spectral observations of 52 disk galaxies 
distributed throughout the cluster field.  No strong emission-line 
galaxies are present within $\sim 400$~kpc of the cluster center, a 
region largely dominated by the similarly-shaped X-ray and low 
surface brightness optical envelopes centered on the giant 
cD galaxy.  However, two-thirds of the galaxies observed outside the 
cluster core exhibit line emission.  \hal\ rotation curves of 14~cluster 
members are used in conjunction with a deep $I$~band image to study the
environmental dependence of the Tully-Fisher relation.  The Tully-Fisher 
zero-point of Abell~2029 matches that of clusters at lower redshifts, 
although we do observe a relatively larger scatter about the Tully-Fisher 
relation.  We do not observe any systematic variation in the
data with projected distance to the cluster center:  we see no environmental 
dependence of Tully-Fisher residuals, $R-I$ color, \hal\ equivalent width,
and the shape and extent of the rotation curves.
\end{abstract}

 
\section {INTRODUCTION}

The evolution of galaxies in clusters is affected by ram-pressure
stripping, tidal interactions, mergers, accretion, and cooling flows.
These processes are expected to be particularly effective in the
richest clusters where they are likely to erase any memory of their
initial conditions (Dressler 1984).  A rich cluster typically has a
conspicuous intracluster medium and a regular,
elliptical-dominated core (Sarazin 1986).  The spiral
galaxies of a rich cluster are predominantly distributed in the periphery
of the cluster, and the closer a spiral disk is to the cluster center, the
less likely it is to contain neutral hydrogen gas (Giovanelli \& Haynes
1985).  Moreover, the frequency and strength of optical emission lines are
lower in cluster galaxies, as first suggested by Osterbrock (1960) and later
verified with large samples of field and cluster galaxies (Gisler 1978;
Dressler et al. 1985; Balogh et al. 1999).  This trend has been
shown to correlate with cluster-centric distance, and is not solely due
to morphological segregation (Balogh et al. 1997).  This lack of
interstellar gas within cluster galaxies may be due to evaporation into the
hotter intracluster gas, or it may be attributed to stripping originating
from either tidal galaxy-galaxy interactions or ram pressure ablation on
intracluster gas.  Ram pressure ablation, which involves the loss of
interstellar gas due to rapid motion through intracluster gas, was first
pointed out by Gunn and Gott (1972) as the likely cause of mass loss of
spiral galaxies in clusters, and optical and 21 cm observations give
direct evidence of this process (Haynes 1990; Kenney \& Koopman 1999).  
In fact, spiral galaxies that
pass through the centers of rich clusters are likely to loose up to 90\% of
their interstellar \HI\ (Roberts \& Haynes 1994).  For example, \HI\
observations of the Virgo cluster and of Abell 2670 (located at a redshift of
$z \sim$0.08 and considerably richer than Virgo) show them to be quite
different.  Indeed, the ``stripping radius'' (the distance from the cluster
center inside which spiral galaxies are \HI\ deficient) is two to
three times larger in Abell 2670 than in the Virgo cluster
(van Gorkom 1996).

Such dramatic environmental effects could affect a variety of observations.
Whitmore, Forbes, \& Rubin (1988) showed that spiral
galaxies within clusters exhibit falling rotation curves, as opposed to the
asymptotically flat or rising rotation curves usually seen in galaxies
located in the periphery of clusters as well as in the field; Adami et
al. (1999) show similar results for late-type spiral galaxies.  Furthermore,
they find that rotation curves of cluster galaxies may be of lower amplitude
than those of field galaxies.  They offer the explanation that the falling
(and lower amplitude) rotation curves are due to mass loss---the inner
galaxies have had their dark matter halos stripped---or that the cluster
environment simply inhibits halo formation.  They also find a monotonic
increase in the mass to light ratio with distance to the cluster center which
they ascribe to the changing shape of the rotation curves with cluster
position.  This view has been contested, however, by Amram et al. (1993) and
Vogt (1995) who find little evidence for any gradients in the outer portions
of rotation curves.  Clusters, the peaks of the density hierarchy, have
undergone strong merger activity, both in terms of large-scale
subclumps (Girardi et al. 1997) and at the galaxy level, as in the
formation of cDs.  In short, there is a large body of work that suggests that
the spiral galaxy population in dense clusters is fundamentally different to
spiral systems found in regions of lower density.

Abell~2029 is one of the densest and richest clusters in the Abell catalog of
rich clusters of galaxies and thus provides an important laboratory in which
to study the effects of the intracluster medium.  It is located at a distance
of \about 240\h\ Mpc (we write the Hubble constant in the form
100$h$ \kms\ Mpc$^{-1}$), and extensive redshift studies have determined its
velocity dispersion to be \about 1500 \kms\ (Dressler 1981; Bower, Ellis, \&
Efstathiou 1988).  In addition, it has been (re)classified as an
Abell richness class 4.4 cluster (Dressler 1978).  The cluster is a textbook
example of a compact, relaxed, cD galaxy-dominated cluster with a high
intracluster X-ray luminosity (1.1 $\times 10^{45} h^{-2}$ ergs s$^{-1}$;
David et al. 1993).  The cD galaxy is one of the largest galaxies known, with
low surface brightness emission detected out to a radius of 0.6\h\ Mpc (Uson,
Boughn, \& Kuhn 1991).

We have obtained a large set of rotation curves of galaxies located in the
field of Abell~2029 in order to study the environmental effects due to the
cluster by comparing our data to the $I$~band Tully-Fisher template relation
for clusters obtained by Giovanelli et al. (1997) and Dale et al. (1999;
hereafter G97 and D99 respectively).  This relation was derived from the
application of the Tully-Fisher relation to more than 1000 galaxies located
in 76 clusters, of which 75 are Abell richness class 2 or lower.  Our
observations are described in Section~2 and the results are presented in
Section~3.  The implications of this work are discussed in Section~4.

\section{The Data}
\label{sec:data}

\subsection{Optical Spectroscopy}  

We obtained long-slit spectroscopy to derive optical rotation curves of
galaxies in Abell~2029.  The observations were carried out at the Mt.~Palomar
5~m telescope during the nights of 1998 April 27--29.  We used the red camera
of the Double Spectrograph (Oke and Gunn 1982) to observe the
\hal\ (6563~\AA), \NII\ (6548, 6584 \AA), and \SII\ (6717, 6731 \AA) emission
lines.  The spatial scale of CCD21 (1024$^2$) was 0$\farcs$468 pixel$^{-1}$.
The combination of the 1200~lines~mm$^{-1}$ grating and a 2\arcsec\ wide slit
yielded a dispersion of 0.65~\AA~pixel$^{-1}$ and a spectral resolution of
1.7~\AA\ (equivalent to 75 \kms\ at 6800 \AA).  The grating angle allowed us
to observe \hal\ in galaxies with recessional velocities between 7,600 and
38,100~\kms.

We were fortunate to enjoy extremely mild atmospheric conditions at
Mt.~Palomar.  All three nights were photometric and dark.  The seeing was
remarkably sharper and more stable than typically encountered at the site; we
estimate the median seeing to have been 1\arcsec, but at times the seeing
dropped to 0$\farcs$6.  Such excellent spatial resolution is important to
obtain high sensitivity rotation curves at the redshifts of the target
galaxies.  Besides yielding higher signal-to-noise per pixel, a sharper
seeing also allows a more accurate placement of the slit.  This is important
because slit offsets and incorrect estimations of the position angles of
galaxy disks can lead to serious errors in the inferred velocity widths
(Bershady 1998, Giovanelli et al. 2000; hereafter G00).  We did not obtain
absolute flux calibrations as they were not necessary for the purpose of this
paper.

We used deep $R$ and $I$ band images to select candidate galaxies as well as
to estimate their position angles.  We discuss these data in the next section.
We observed all probable disk-like systems on the reference images that might
be members of the cluster, did not appear to be face-on, and were free of
contamination from foreground stars.  The limited resolution of the reference
images precluded unambiguous identification of appropriate Tully-Fisher
candidates.  Our observing strategy began with a five minute test-exposure
on each spectroscopic target.  That way we were able to estimate ``on the
fly'' the exposure time required in order to sample adequately the outer disk
regions.  Furthermore, the test exposure determined whether the galaxy was
even useful to our work; a galaxy may lie in the foreground or background of
the cluster or it may contain little or no \hal\ emission.  If the
observation was deemed useful, a second exposure typically ranged between
15 and 45~minutes.  We detected line emission in half of the 52~observed
galaxies.  We list the galaxies observed in Table~1, sorting the entries by
Right Ascension.  The table contains:

\noindent 
Col. 1: Identification names corresponding to a coding number in our database,
referred to as the Arecibo General Catalog.

\noindent
Cols. 2 and 3: Right Ascension and Declination in the 1950.0 epoch.
Coordinates have been obtained from the Digitized Sky Survey catalog and are
accurate to $<$ 2\arcsec.

\noindent
Col. 4: The galaxy radial velocity as measured in the heliocentric reference
frame.  The redshift measurements of the galaxies without emission lines
were obtained from the NED\footnote{The NASA/IPAC Extragalactic Database is
operated by the Jet Propulsion Laboratory, California Institute of Technology,
under contract with the National Aeronautics and Space Administration.}
database.  They have been previously derived by others using absorption-line
spectra.

\noindent
Col. 5: An indication of the usefulness of the optical emission lines in
order to apply the Tully-Fisher relation: 0=no lines present; 1=strong
emission lines throughout much of the disk; 2=weak or nuclear emission only.

Rotation curves are extracted as discussed in Dale et al. (1997 and 1998;
hereafter D97 and D98).  We use the \hal\ emission line to map the rotation
curve except in the case of the galaxy AGC 251909 where the emission of the
\NII\ line (6584~\AA) extends to a larger distance than that of the
\hal\ emission.  We center the rotation curve kinematically  by assigning the
velocity nearest to the average of the 10\% and 90\% velocities to be at
radius zero, where an N\% velocity is greater than N\% of the velocity data
points in the rotation curve.  The average of the 10\% and 90\% velocities is
taken to be the galaxy's recessional velocity.  We define the observed
rotational velocity width to be \Wobs\ $\equiv V_{\rm 90\%} - V_{\rm 10\%}$.
We filled-in small portions of the \hal\ rotation curve of two galaxies
(AGC 251913 and AGC 251912) using data from the \NII\ rotation curve in order
to provide information on the shape of the inner parts and to ensure
consistent estimates of \Wobs.

The rotation curves in our sample vary in physical extent, and more
importantly, they do not all reach the optical radius, \Ropt, the distance
along the major axis to the isophote containing 83\% of the $I$~band flux.
This radius is reported by Persic \& Salucci (1991)
and G00 to be the most useful radius at which to measure the velocity width
of rotation curves.  We have extrapolated the rotation curves, and hence made
adjustments to \Wobs, when they did not reach \Ropt.  The resulting
correction, ${\Delta}_{\rm sh}$, depends on the shape of the rotation curve
and only exceeded 4\% for AGC 251831 where the correction was large
($\sim 44$\%).

To recover the actual velocity widths, a few more corrections are necessary.
The first is the factor 1/sin$i$ to convert the width observed when a disk is
inclined to the line of sight at an angle $i$ to what would be observed if
the disk were edge--on, and the second is the factor 1/(1+$z$) to correct
the cosmological broadening of $W$.  A final correction,
$f_{\rm slit} < 1.05$, accounts for the finite width of the slit of the
spectrograph (G00).  The corrected optical rotational velocity width is

\be
W_{\rm cor} = {{W_{\rm obs} + {\Delta}_{\rm sh}} \over {(1+z)\sin i}}
f_{\rm slit}.
\ee
A discussion of the errors in the velocity widths can be found in D97.

\begin{figure}[ht]
\centerline{\psfig{figure=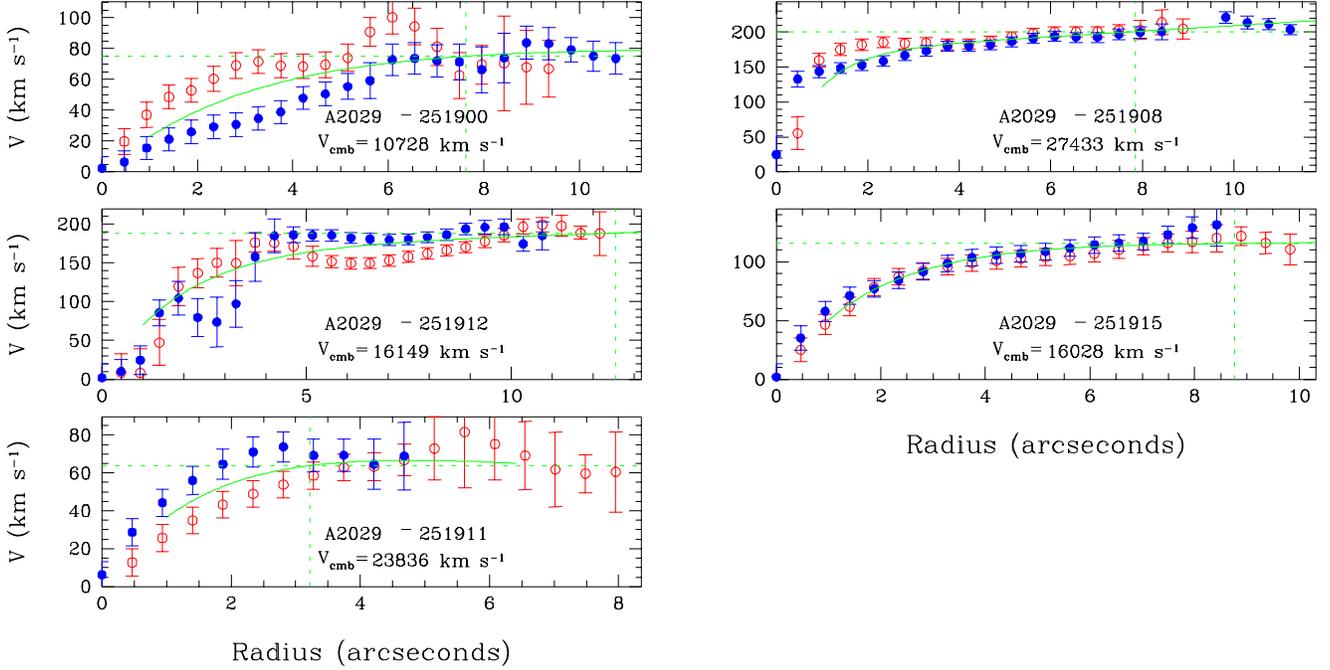,width=7.0in,bbllx=67pt,bblly=485pt,bburx=515pt,bbury=714pt}}
\caption[]
{\ The kinematically folded rotation curves (see text).  The error bars
include both the uncertainty in the wavelength calibration and the rotation
curve fitting routine used.  Names of the galaxy are given along with the
CMB radial velocities.  Two dashed lines are drawn: the horizontal line
indicates the adopted half velocity width, $W$/2, which in some cases arises
from an extrapolation to the RC; the vertical line is at \Ropt, the radius
containing 83\% of the $I$ band flux.  A fit to the rotation curve is
indicated by a solid line.  Note that the rotation curves are {\it not}
deprojected to an edge-on orientation.}
\label{fig:RCs}
\end{figure}

Figure \ref{fig:RCs} is a display of the rotation curves observed in the
field of Abell~2029.  Entries in the figure are sorted by Right Ascension.
The name of the galaxy is given along with the CMB radial velocity.  Two
dashed lines are drawn: the vertical line is at \Ropt; the horizontal line
indicates the adopted half-velocity width, $W$/2, which in some cases arises
from an extrapolation to the rotation curve (see Table 1).  Overlayed are
the fits used to infer $W$(\Ropt).  Details of the fitting procedure can be
found in G00.  The error bars include both the uncertainty in the wavelength
calibration and the routine used to fit the rotation curve.  Notice that the
data are highly correlated due to seeing and guiding jitter.  This is
properly taken into account by the fitting routines (see D97 and references
therein for details).
%%%%%%%%%%%%%%%%%%%%%%%%%%%%%%%%%%%%%%
\centerline{\psfig{figure=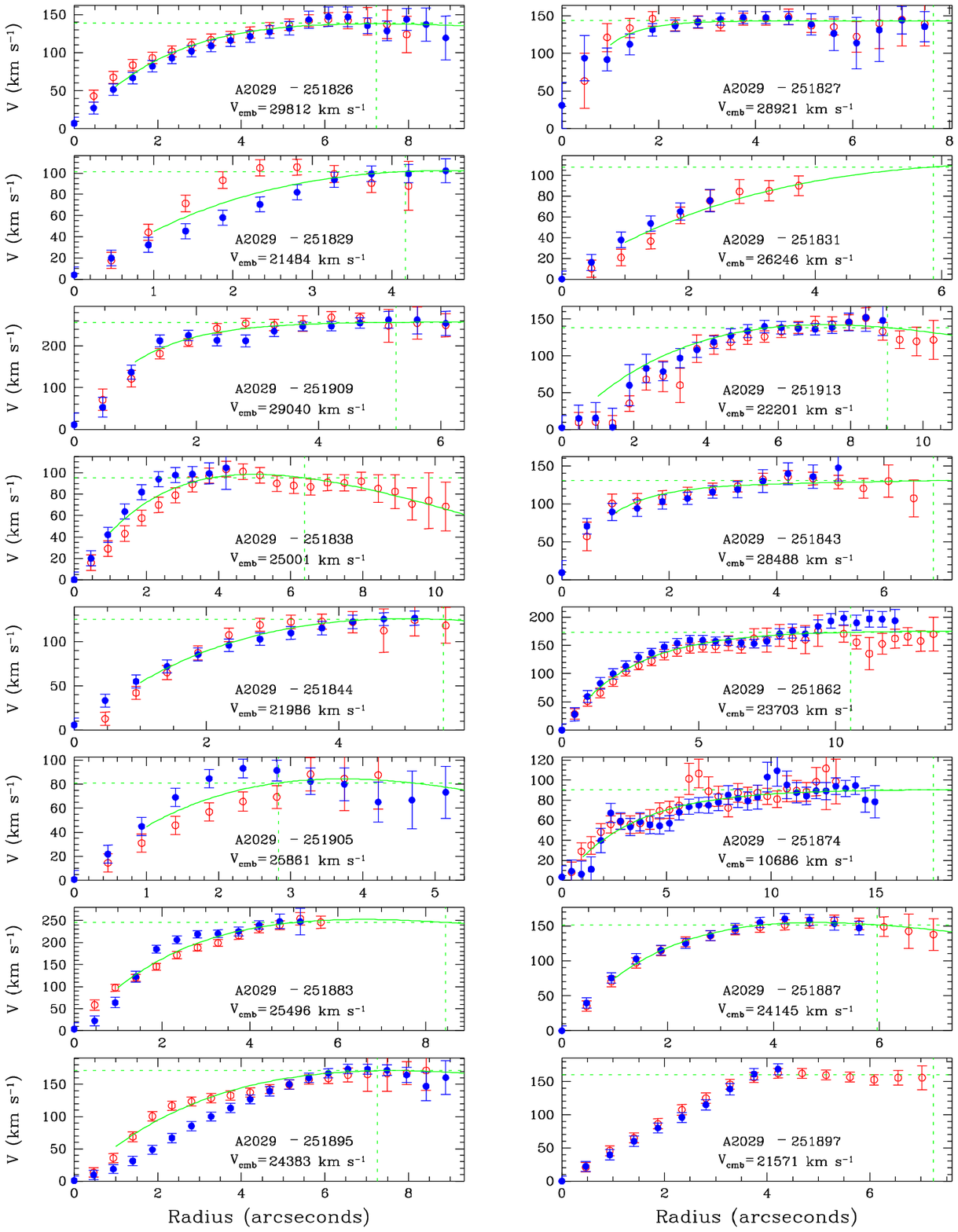,width=7.0in,bbllx=67pt,bblly=144pt,bburx=515pt,bbury=714pt}}
%\newpage
%%%%%%%%%%%%%%%%%%%%%%%%%%%%%%%%%%%

We list in Table~2 the complete set of spectroscopic data corresponding to the
21~galaxies for which we obtained useful rotation curves, sorting the entries
by Right Ascension.  The parameters listed in Table~2 are:

\noindent 
Col. 1: Identification names corresponding to a coding number in the Arecibo
General Catalog.

\noindent
Col. 2: The spectral exposure time in seconds.

\noindent
Col. 3: The recessional velocity of the galaxy in the CMB reference frame,
assuming a Sun-CMB relative velocity of 369.5 \kms\ towards
 $(l,b)=264.4^\circ,48.4^\circ)$ (Kogut et al. 1993).  Errors are
parenthesized: e.g. 13241(08) means 13241$\pm$08.

\noindent
Col. 4: The observed velocity width in \kms.

\noindent
Col. 5: The velocity width in \kms\ at \Ropt\ after correcting for the shape
of the rotation curve, the cosmological broadening, and the smearing effects
due to the finite width of the slit of the spectrograph.

\noindent
Col. 6: The corrected velocity width in \kms\ converted to an edge-on
perspective.

\noindent
Col. 7: The adopted inclination of the plane of the disk to the line of
sight, $i$, expressed in degrees, (90$^\circ$ corresponds to an edge--on
perspective); the derivation of $i$ and its associated uncertainty are
discussed in Section~4 of D97.

\noindent
Col. 8: The logarithm in base~10 of the corrected velocity width (value in
column 6), together with its estimated uncertainty in parentheses.  The 
uncertainty takes into account both measurement errors and uncertainties
arising from the corrections.  The format 2.576(22), for example, is
equivalent to 2.576$\pm$0.022.

\subsection{Optical Imaging}

$I$~band photometry of Abell~2029 was obtained for a different project by one
of us (JMU) in collaboration with S. P. Boughn (Haverford) with the 0.9~m
telescope on Kitt Peak National Observatory on 1998~April~19.  They used the
T2KA camera mounted at the f:7.5 Cassegrain focus which resulted in square
pixels, 0$\farcs$68 on a side.  The seeing was excellent, between 0$\farcs$7
and 0$\farcs$9, which resulted in an effective seeing of $\sim1\farcs$2 due
to the available pixel size.

Two sets of fifteen partially overlapping frames were used to form a mosaic
of about 35$\arcmin$ (RA) by 58$\arcmin$ (Dec).  The central three by three
mosaic has overlaps of about 3/4 of a frame between immediately adjacent
frames, whereas the outer three frames to the north and south overlap by about
1/2 of a frame with the closest of the central ones.  All but four of the
outlying frames to the north and south were obtained with air masses between
1.11 and 1.25.  The exposures lasted five minutes.  The data were processed
as discussed in Uson, Boughn and Kuhn (1991, hereafter UBK91).  All frames
were used to generate a ``sky-flat'' gain calibration frame.  Since the
cluster contains a diffuse halo that surrounds the central galaxy, a
12$\arcmin$ by 12$\arcmin$ area centered on the cluster was blanked on all
the relevant frames before using them to generate the sky flat as discussed
in UBK91.  The calibrated frames were used to determine the secant-law
extinction which had a slope of 0.06 mag/airmass.

Absolute calibration was done using stars from Landolt's $UBVRI$ secondary
calibration list (Landolt 1983).  Details will be given elsewhere.

The $R$~band photometry was obtained from UBK91.

Flux estimation follows from the data reduction methods discussed in D97 and
D98 using both standard and customized IRAF\footnote{IRAF (Image Reduction
and Analysis Facility) is distributed by NOAO.} packages.  We will only
mention here that the measured fluxes, denoted $m_{\rm obs}$, include
extrapolations of the exponential fits to the surface brightness profiles to
eight disk scale lengths and are typically accurate to \about\ 0.03 mag
(uncertainties at least as large are later included after making corrections
for internal extinction).  We apply some corrections to $m_{\rm obs}$ in
order to obtain the final $I$ band fluxes:
\be
m_I = m_{\rm obs} - A_I + k_I - \Delta m_{\rm int}.
\ee
For the Galactic extinction correction $A_I$, we use the recent work of
Schlegel, Finkbeiner, and Davis (1998) who have provided accurate Galactic
reddening estimates using COBE/DIRBE and IRAS/ISSA dust maps.  The internal
extinction correction, $\Delta m_{\rm int}$, is applied using the procedure
outlined in G97,
\be
 \Delta m_{\rm int} = - f(T) \; \gamma(W_{\rm cor}) \; \log(1-e),
\ee
where $\gamma$ ($\lesssim$ 1.0) depends on the corrected velocity width
$W_{\rm cor}$ and $e$ is the ellipticity of the spiral disk, corrected
for atmospheric seeing effects as described in Section 5 of D97 (the adopted
correction $\Delta m_{\rm int}$ is slightly smaller for early, less dusty
galaxies: $f(T)$=0.85 for types $T$ earlier than Sbc; $f(T)$=1 otherwise).
We apply a cosmological k-correction according to Han (1992):
$k_I = (0.5876 - 0.1658$T$)z$.

The relevant photometric data are listed in Table~3 with the first column
matching that of Table~2.  The remaining parameters are:

\noindent
Col. 2: Morphological type code in the RC3 scheme, where code 1 corresponds 
to Sa's, code 3 to Sb's, code 5 to Sc's and so on.  We assign these
codes after visually inspecting the CCD $I$ band images and after noting the
value of $R_{\rm 75}/R_{\rm 25}$, where $R_X$ is the radius containing
X\% of the $I$ band flux.  This ratio is a measure of the central
concentration of the flux which was computed for a variety of
bulge--to--disk ratios.  Given the limited resolution of the images, some
of the inferred types are rather uncertain; uncertain types are followed
by a colon.

\noindent
Col. 3: The angular distance $\theta$ in arcminutes from the center of each 
cluster.

\noindent
Col. 4: Position angle used for spectrograph slit positioning (North:
0$^{\circ}$, East: 90$^{\circ}$).

\noindent
Col. 5: Ellipticity of the disk corrected for seeing effects as described in
Sec. 5 of D97, along with its corresponding uncertainty expressed using the
same convention as in Table~2.
 
\noindent
Col. 6: The (exponential) disk scale length in arcseconds.

\noindent
Col. 7: The distance along the major axis to the isophote containing 83\% of
the $I$ band flux.

\noindent
Col. 8: The measured $I$ band magnitude, extrapolated to 8 disk scale lengths
assuming that the surface brightness profile of the disk is well described
by an exponential function.

\noindent
Col. 9: The absolute magnitude, computed assuming that the galaxy is at the 
distance indicated by the cluster redshift, or by assuming the galaxy is at
the distance indicated by the redshift if the galaxy is not deemed to be a
member of the cluster.  The calculation assumes
$H_\circ = 100h$ \kms\ Mpc$^{-1}$, so the value listed is strictly
$M_I - 5\log h$.  This parameter is calculated after expressing the redshift
in the CMB frame and neglecting any peculiar motion.  The uncertainty on the
magnitude, parenthetically included in hundredths of a mag, is the sum in
quadrature of the measurement errors and the estimate of the uncertainty in
the corrections applied to the measured parameter.

\noindent
Col. 10: The difference in the $R$ and $I$ band magnitudes.

When an asterisk appears at the end of the line, we include a detailed
comment on that particular object.  Because of the extent of these comments
we have not appended them to the table but have included them in the text.
Note that a record is flagged in both Tables~2 and~3, whether the comments
refer to the photometry, to the spectroscopy, or to both.
  
\small
\noindent 251826: Background galaxy.\\
\noindent 251827: Background galaxy; uncertain PA.\\
%\noindent 251829: \\
\noindent 251831: Rising rotation curve; large rotation curve extrapolation.\\
\noindent 251909: Background galaxy; \NII\ rotation curve used.\\
\noindent 251913: \NII\ patch for radii $<$ 3\arcsec; uncertain disk
ellipticity.\\
\noindent 251838: Asymmetric $I$ band profile.\\
\noindent 251843: Note low $i$.\\
%\noindent 251844: \\
%\noindent 251862: \\
%\noindent 251905: Very small galaxy.\\
\noindent 251874: Foreground galaxy; 5 minute integration.\\
%\noindent 251883: \\
%\noindent 251887: \\
\noindent 251895: Uncertain disk ellipticity; asymmetric $I$ band profile.\\
\noindent 251897: Center-of-light used for rotation curve spatial and
kinematic center.\\
\noindent 251900: Foreground galaxy; 5 minute integration; center-of-light
used to determine the center of the rotation curve.\\
\noindent 251908: Tidally interacting with small companion 13\arcsec\ to NE
(\S \ref{sec:tidal}).\\
\noindent 251912: Foreground galaxy; \NII\ patch for radii $<$ 3\arcsec;
flux disentanglement with AGC 251911 difficult.\\
\noindent 251911: Flux disentanglement with AGC 251912 difficult; uncertain
PA and disk ellipticity.\\
\noindent 251915: Foreground galaxy.\\
\normalsize

\section{Results}

\subsection{The Distribution of Emission-Line Galaxies}

\begin{figure}[ht]
\centerline{\psfig{figure=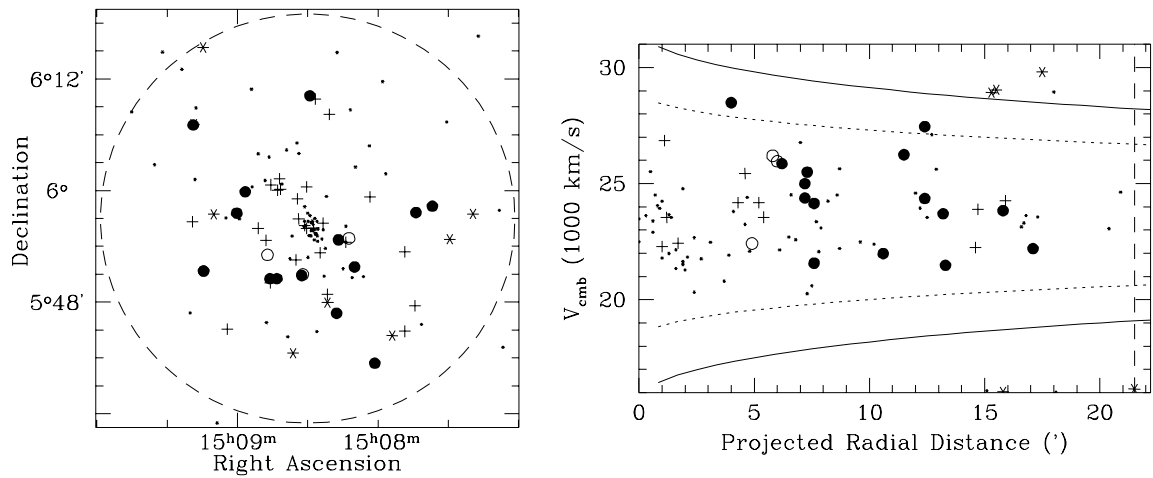,width=6.0in,bbllx=20pt,bblly=363pt,bburx=591pt,bbury=600pt}}
\caption[Cluster Membership]
{\ Sky and velocity distribution of galaxies in the cluster field.  Circles
represent cluster members with measured photometry and widths; if unfilled,
widths are poorly determined.  Asterisks identify foreground and background
galaxies, and dots give the location of galaxies with known redshift.
Crosses indicate the positions of galaxies lacking emission lines.  The
dashed line in each panel indicates 1~Abell radius.  The dotted and solid
lines respectively indicate 2~and 3$\sigma$ cluster membership contours
(Carlberg et al. 1997).}
\label{fig:field}
\end{figure}

A plot of the sky distribution of Abell~2029 field objects is displayed in
the left panel of Figure \ref{fig:field}.
%%%%%%%%%%%%%%%%%%%%%%%%%%%%%%%%%%%%%%
%%%%%%%%%%%%%%%%%%%%%%%%%%%%%%%%%%%
Filled (open) circles represent galaxies with fair to strong (weak but
detectable) optical emission lines.  Asterisks mark foreground and background
galaxies with observed rotation curves and crosses indicate the positions of
observed galaxies lacking emission lines.  Other galaxies with known
redshifts are indicated by small dots.  The right panel in
Figure~\ref{fig:field} presents galaxy redshifts versus their projected
distances from the cluster center.  In both panels one Abell radius
(1.48\h\ Mpc) is indicated by the dashed line.  The redshift measurements
of the galaxies lacking emission lines (indicated by crosses) are drawn from
the literature.

All the galaxies we observed spectroscopically are projected to lie within
one Abell radius, but of course the ultimate definition of cluster membership
relies on the observed redshift distribution.  Using spectra for 47~cluster
galaxies, Dressler (1981) found a velocity dispersion of 1430 \kms\ for the
cluster.  This was confirmed by Bower et al.~(1988) for both the inner and
outer regions of the cluster using a similar sample size.  Combining our
observations with those reported in the literature yields a redshift sample
twice as large ($N=99$) as those used previously.  We find a mean CMB cluster
velocity of 23,657$\pm^{216}_{99}$ \kms\ and a 1$\sigma$ (rest frame)
velocity dispersion of 1454~\kms.  The cluster systemic velocity agrees well
with the CMB redshift velocity of the cD galaxy: 23,550 \kms\ (de Vaucouleurs
et al.~1991).
This is not surprising given that the cD galaxy is about 100~times more
luminous than any other cluster member (UBK91).  We derive (projected)
cluster membership contours using the results from the CNOC survey of galaxy
clusters (Carlberg et al.~1997), scaled by our estimate of the cluster
velocity dispersion for Abell 2029.  The 3$\sigma$ contour is indicated by
the two solid curves in the right-hand panel and, coupled with the Abell
radius of 21$\farcm$5, represents our working definition of the cluster
proper; the dotted curves show the 2$\sigma$ contour (see, for example,
Balogh et al.~1999).

The most striking aspect of Figure~\ref{fig:field} is the paucity of observed
emission line galaxies located in the cluster core.  We observed 17~of the
galaxies that are projected to lie within the inner 5$\arcmin$
(\about 340\h\ kpc) of the cluster, and the only one exhibiting a strong
emission line is outside the 2$\sigma$ cluster-membership contour.  Of the
remaining 16~inner-cluster galaxies observed, one has a weak \hal\ line, but
the other 15~galaxies show no emission lines at all in a 300~second test
observation.  Our success rate was much higher when observing galaxies that
are projected to lie further than 5$\arcmin$ from the cluster center.
Indeed, 24~of the 35~galaxies observed in that area had emission lines that
were detected in the 5~minute test exposure.

\begin{figure}[ht]
\centerline{\psfig{figure=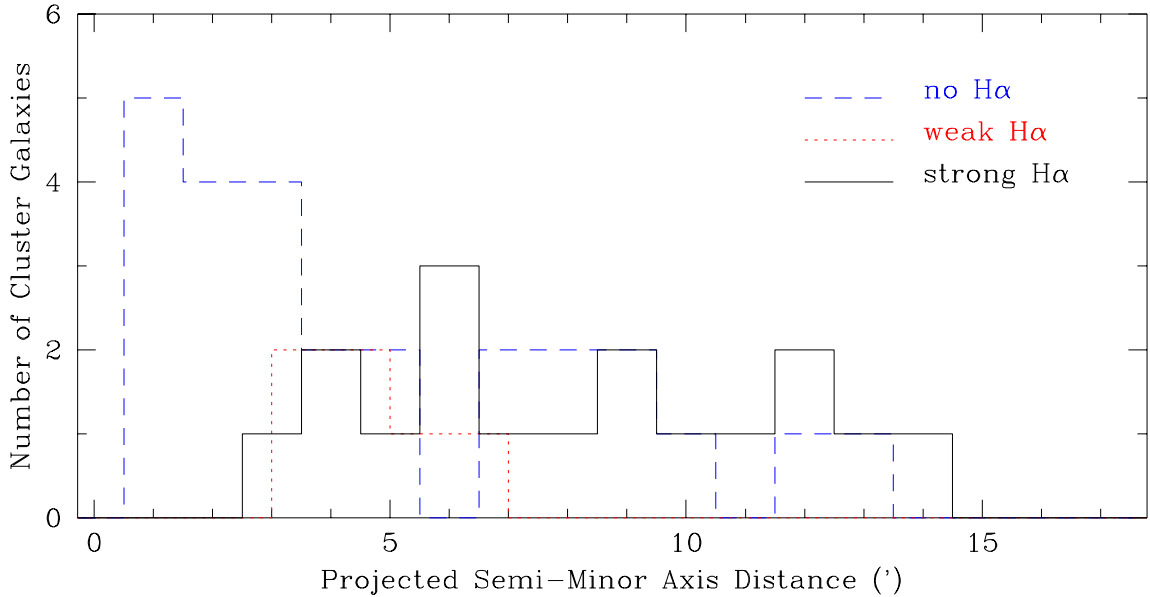,width=6.0in,bbllx=46pt,bblly=430pt,bburx=564pt,bbury=700pt}} \caption[Cluster Membership] {\ The spatial distribution of observed cluster galaxies as a function of the projected distance from the cD galaxy's semi-minor axis, separated according to the strength of the observed \hal\ line emission.  The counts
are computed from within elliptical bins of constant axis ratio 2:1 and
oriented at a position angle 21$^\circ$ East of North.}
\label{fig:histogram}
\end{figure}

An alternative display of such data is presented in
Figure \ref{fig:histogram}.  Here we show the relative spatial
distribution of the galaxies exhibiting strong, weak, and no \hal\ emission
after accounting for the non-circular orientation of the projected diffuse
optical light of the cD galaxy.  The counts are computed in
elliptical bins of constant axis ratio 2:1, centered on the cD and oriented at
a position angle 21$^\circ$ East of North (UBK91); the X-ray
brightness contours are consistent with this morphology (Slezak, Durret
\& Gerbal 1994).
%%%%%%%%%%%%%%%%%%%%%%%%%%%%%%%%%%%%%%
%%%%%%%%%%%%%%%%%%%%%%%%%%%%%%%%%%%
Again, there is a fairly clear delineation between regions with many
\hal--rich galaxies and regions without \hal--emitting galaxies.  For
comparison purposes we note that UBK91 found  that approximately 90\% of the
projected $R$~band emission of the cD~galaxy falls within an ellipse with a
semi-minor axis of about 4\arcmin.

\subsection{Tully-Fisher Data}
\begin{figure}[ht]
\centerline{\psfig{figure=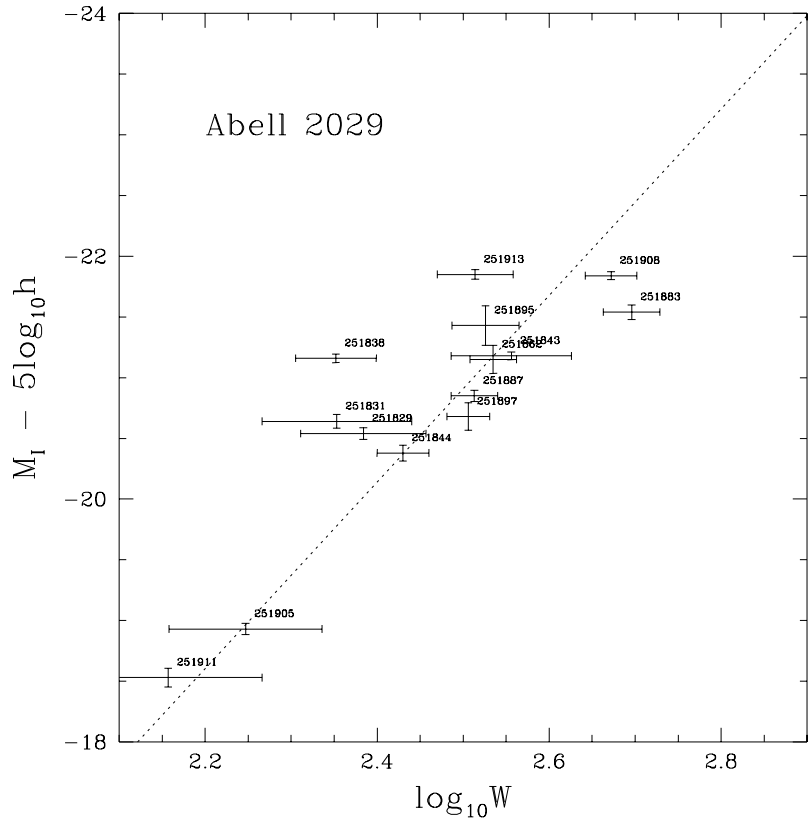,width=5.0in,bbllx=18pt,bblly=157pt,bburx=567pt,bbury=693pt}}
\caption[Cluster Membership]
{\ The Tully-Fisher data for Abell~2029, uncorrected for cluster peculiar
motion.  The dashed line is the template relation for clusters obtained by
D99, Eqn.~\ref{eq:TF}.}
\label{fig:TF}
\end{figure}

The Tully-Fisher data for all members of Abell~2029 are presented in
Figure~\ref{fig:TF}.
%%%%%%%%%%%%%%%%%%%%%%%%%%%%%%%%%%%%%%
%%%%%%%%%%%%%%%%%%%%%%%%%%%%%%%%%%%
Included in the Tully-Fisher plot is the template relation obtained from
the cluster Tully-Fisher study of D99:
\be
y = -7.68x - 20.91
\label{eq:TF}
\ee
where $y$ is $M_I - 5\log h$ and $x$ is log$W_{\rm cor} - 2.5$.  The data are
corrected for the effects described in Section \ref{sec:data}.  In addition,
the morphological type offsets for early-type disk galaxies advocated by G97
and D97 are applied to three galaxies: $\Delta m_T=-0.1$ mag for the Sb
galaxy AGC~251913, and $\Delta m_T=-0.32$ mag for AGC~251862 (Sa) and
AGC~251912 (S0/a).  The scatter in the Tully-Fisher data is 0.56 mag,
considerably larger than the values of 0.35 and 0.38 mag found in G97
and D99, respectively.  Notice the absence of a detectable offset between the
Abell~2029 data and the Tully-Fisher template.  Although we have not included
any corrections for Malmquist bias, typically of order 0.04 magnitudes for
clusters with 10 or more Tully-Fisher measurements (D99), it gives us great
confidence in our data reduction procedures as the $I$~band photometry was
reduced independently of this project.

\subsection{Are There Other Trends with Cluster Environment?}

\begin{figure}[ht]
\centerline{\psfig{figure=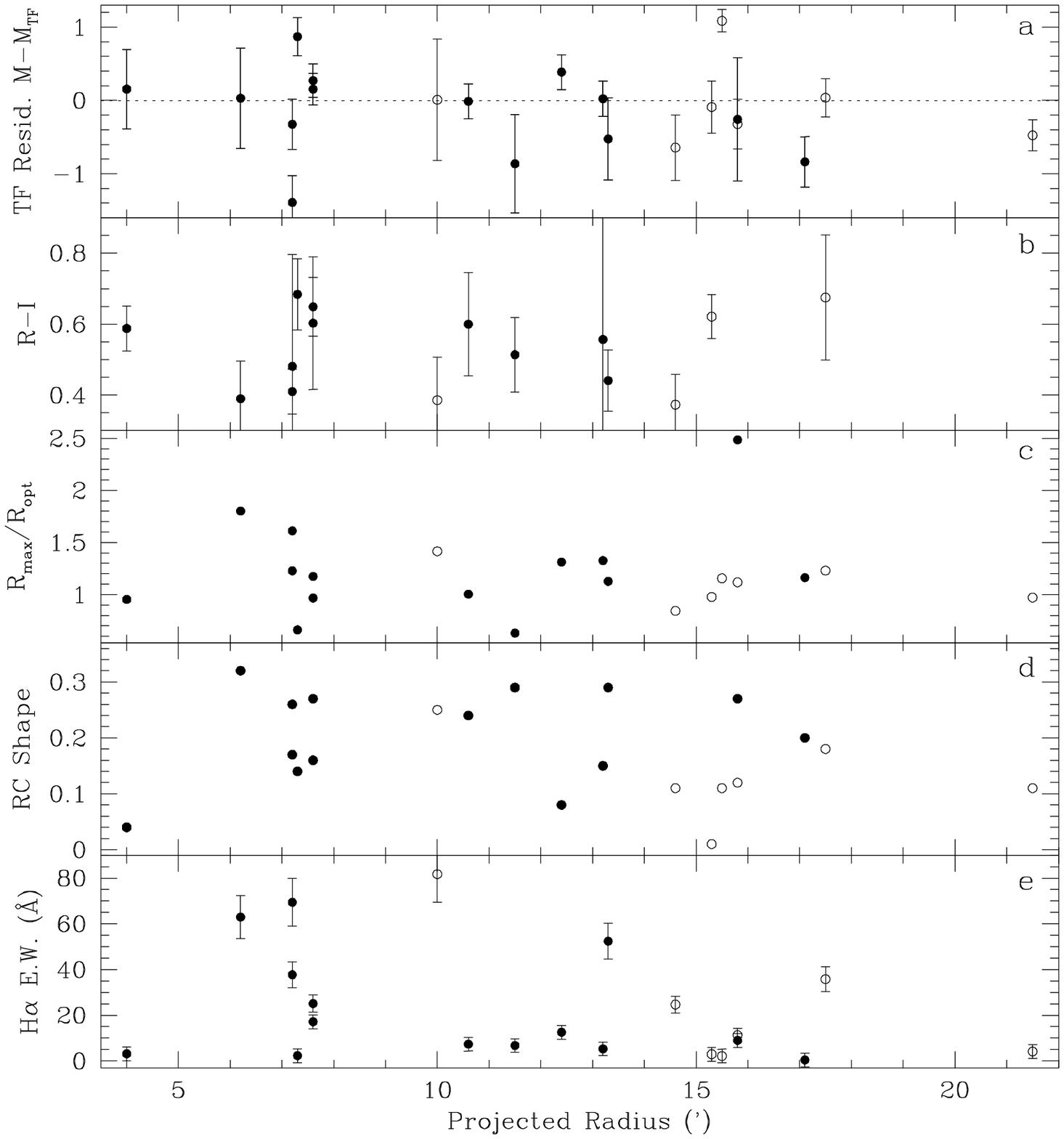,width=5in,bbllx=27pt,bblly=143pt,bburx=565pt,bbury=724pt}}
\caption[Cluster Membership]
{\ Some properties of the emission-line galaxies plotted versus projected
distance to the center of the cluster.  Filled (open) circles represent
cluster (foreground and background) galaxies.
Panel (a) shows the residuals of the Tully-Fisher data where the error bars
derive from a quadrature sum of the errors in the magnitudes, velocity widths,
and the template zero-point.  Panels (b) and (c) give the $R-I$ colors and
the extent of the \hal\ rotation curve (normalized by the optical radius),
respectively.  Panel (d) displays the outer gradient, or shape, of the
rotation curve, and Panel (e) plots the \hal\ equivalent widths.}
\label{fig:TFresiduals}
\end{figure}

Plotted in Figure \ref{fig:TFresiduals} are some properties of the
emission-line galaxies as a function of projected distance from the cluster
center.  Filled (open) circles represent cluster (foreground and background)
galaxies.
%%%%%%%%%%%%%%%%%%%%%%%%%%%%%%%%%%%%%%
%%%%%%%%%%%%%%%%%%%%%%%%%%%%%%%%%%%
The top panel displays the residuals of the Tully-Fisher data.  An overly
large $M/L$ ratio would correspond to an overly fast rotating disk for a
given absolute magnitude, i.e. the galaxy would appear fainter than the
fiducial template would indicate and hence would have a positive residual.
An increasing $M/L$ ratio with increasing projected distance, as advocated
by Whitmore, Forbes \& Rubin~(1988) and Adami et al.~(1999), would thus
appear as a trend of increasing residuals with increasing projected distance.  Such a trend is not present in the data.  In fact, the slight trend seen for cluster galaxies acts in the opposite sense.

The data in Panels b and~c show that no trend in $R-I$ color or in the
physical extent of the rotation curve with distance between the galaxy and
the center of the cluster ($R_{\rm max}$ is the maximum radial distance at
which the rotation curve is measured).  The shape of the rotation curve
(Panel d), and the \hal\ equivalent width (Panel e) seem uncorrelated as well
with location in the cluster.  The ``shape'' of a rotation curve is defined
here as the outer gradient of the rotation curve, namely
\be
{\rm Rotation \;\; Curve \;\;Shape} = {W(R_{\rm opt}) - W(0.5R_{\rm opt})
\over W(R_{\rm opt})}.
\ee

\subsection{A Tidal \hal\ Bridge Between AGC 251908 and its Companion}
\label{sec:tidal}

\begin{figure}[ht]
\centerline{\psfig{figure=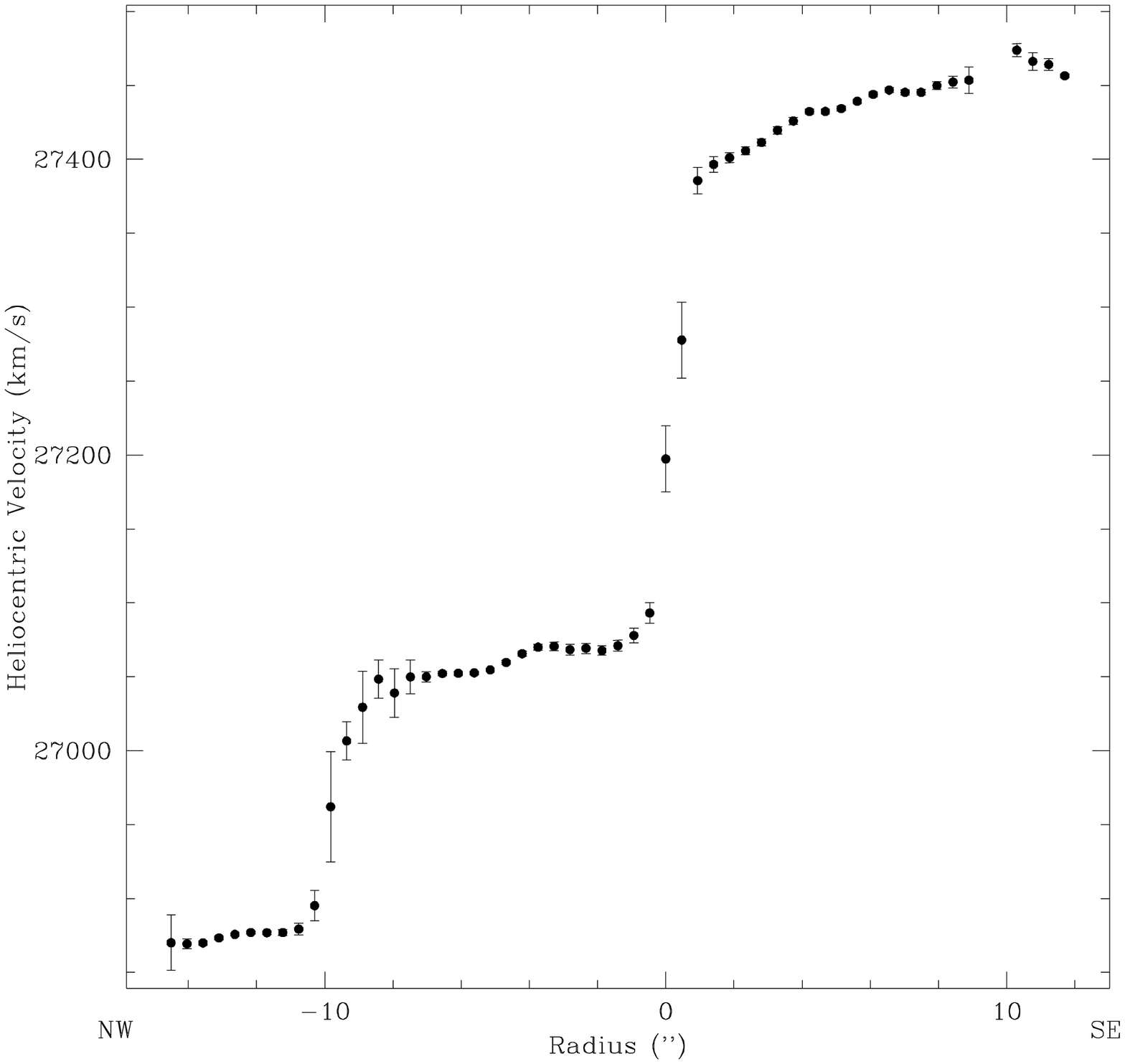,width=4.0in,bbllx=18pt,bblly=157pt,bburx=567pt,bbury=693pt}}
\caption[Cluster Membership]
{\ The observed rotation curve for AGC 251908 and its tidally-connected
companion.  No rotation is seen in the companion galaxy due in part to the
discrepancy in the position angle of its ``disk'' and that used for the
long-slit spectroscopic observation.}
\label{fig:tidal}
\end{figure}

One reason why we took spectra of all plausible disk systems in the cluster
core was to try to observe the tidal stripping of a galaxy passing through a
dense intracluster medium.  For example, AGC 250201 is a possible candidate
as it lies well within the low surface brightness envelope of the cD galaxy
but has a redshift more than 3000 \kms\ higher than the cD.  Unfortunately,
no core galaxies show strong emission lines as discussed above.  We did
observe, however, a bridge of \hal\ emission connecting one pair of galaxies
outside the cluster core: AGC 251908 and its small satellite/companion galaxy
located approximately 13\arcsec\ to the NW.  The joint rotation curve
for this pair is displayed in Figure~\ref{fig:tidal}.
%%%%%%%%%%%%%%%%%%%%%%%%%%%%%%%%%%%%%%
%%%%%%%%%%%%%%%%%%%%%%%%%%%%%%%%%%%
The companion is amorphous in $I$ band and has a FWHM of 1$\farcs$7.  The
flux in a circular aperture of radius 2$\farcs$7 centered on the companion
corresponds to a magnitude of $m_I=19.87$, or $M_I=-17.08$.  No such bridge
is evident in the $I$ band image, nor is any obvious warping seen in the disk
of AGC 251908.

This region of the sky was unfortunately not covered by our $R$ band map, so
we turn to the Palomar Observatory Sky Survey II\footnote{The Second Palomar
Observatory Sky Survey was made by the California Institute of Technology
with funds from the National Science Foundation, the National Geographic
Society, the Sloan Foundation, the Samuel Oschin Foundation, and the Eastman
Kodak Corporation.} data (1\arcsec\ pixels) for further clues.  We notice
that the companion has a different morphology on the POSS-II red plate, the
plate whose filter bandpass includes the \hal\ line.  The emission from the
companion is significantly more extended (3\farcs4 FWHM) at this wavelength,
and it clearly has an elongated shape.  It is reasonable to assume that the
companion is a disk galaxy since we see \hal\ in emission.  Moreover, we
estimate the position angle of the major axis of the disk to be 45$^\circ$,
essentially perpendicular to the angle used for the long-slit observation of
AGC~251908.  Thus it is not surprising that we detected little rotation of
the companion galaxy.

\section{Discussion}

The extended optical envelope centered on the cD galaxy
in Abell 2029 was interpreted by UBK91 as the leftover of a
period of violent relaxation during the initial cluster collapse.
This process would have led as well to the ionization of the gas present in 
the galaxies that participated in such a collapse and to the hot intracluster 
gas, observable through its X-ray emission and strongly peaked at the 
cluster center.  The galaxies that traverse the cluster core will have their 
interstellar gas stripped by the dense intracluster gas, and is 
likely the root cause of the
lack of emission line signatures from these galaxies.  Indeed, UBK91 show
the extremely large extent of the $R$~band profile which they trace to \about
5$\arcmin$, a boundary similar to the delineation between emission-poor and
relatively emission-rich environs.  Beyond this distance, however, it appears
that the projected distance from the cluster core at which a cluster galaxy
lies plays only a small role in the \hal\ success rate: galaxies with emission
lines are spatially evenly mixed with those galaxies lacking emission lines.

A goal of this project was to search for signs of tidal stripping on
inner-core galaxies.  Therefore, we observed all disk systems in the cluster
core, irrespective of whether or not the galaxy appeared to be a useful
``Tully-Fisher'' galaxy.  This could mean that we inspected a
large fraction of early type spirals (S0 to Sab) in the cluster core.  In the
Coma cluster, for example, S0 galaxies outnumber spirals in the cluster center
by more than a factor of two (Andreon \& Davoust 1997).  We find that the
average (RC3) morphological type for the emission line galaxies that we
observed in Abell~2029 is $T=4.7$ (Sbc/Sc),  whereas it is $T=3.2$ (Sb/Sbc)
for the non-emission line
galaxies.   There is a small difference in the average morphological type of
these  two groups, and thus the discrepancy in the spatial distribution of 
emission line galaxies may not be quite so surprising, as early-type disk 
galaxies contain fewer \HII\ regions than are typically found in late-type 
spirals.

% The next sentence sounds very awkward.  I believe it is not necessary and
%should be left out.  Otherwise, we should rewrite it and the previous one
%as well.
%  {\bf However, the discrepancy is not simply that we observed elliptical 
% and early-type spirals in the cluster core and late-type spirals outside
% the core}.

Outside the cluster core, we find no trends in $R-I$ color, \hal\ equivalent
width, the shape and physical extent of a rotation curve, nor Tully-Fisher
characteristics as a function of projected distance from the cluster center.
The latter result is in agreement with the Tully-Fisher work of Biviano
et al.~(1990) and G97 at relatively low redshifts, as well as with the work
at $0.2<z<0.6$ described in D99.

The dispersion in the Tully-Fisher plot for Abell~2029 is about 50\%
larger than what is
seen in clusters at lower redshifts (0.56 mag vs. $0.35-0.38$ mag).
The dispersion does not significantly change if each galaxy is placed
at the distance indicated by its individual redshift, rather than assuming
that all cluster galaxies are at the same distance.
Because of the relative difficulty in assessing such galactic properties
as disk inclination and major axis position angle, such an increase 
is not surprising given the larger measurement uncertainties. 
However, it is also possible that the strong intracluster medium environment of
Abell~2029 promotes a wider variety of $M/L$ ratios than what is typically
found in other clusters.  In other words, rich clusters of galaxies may have
a higher intrinsic contribution to the overall Tully-Fisher scatter.  Using
their sample of 522 late-type galaxies in 52 lower redshift Abell clusters,
D99 find an intrinsic scatter of 0.25 magnitudes; measurement uncertainties
account for the remaining 0.28 magnitudes of the overall scatter.  Similar
numbers have been found for the Tully-Fisher samples of G97 and
Willick~(1999).  In Abell 2029, we find that measurement uncertainties are
responsible for 0.40 magnitudes of the scatter, and thus we infer an
intrinsic scatter of 0.39 magnitudes.  In short, the relatively large
Tully-Fisher dispersion for Abell 2029 appears to be a consequence of both
increased measurement uncertainties and relatively larger variations in the
spiral galaxy population.

Finally, the overall offset of the Abell~2029 Tully-Fisher data, with
respect to the fiducial template relation for lower $z$ clusters, is
indistinguishable from a null offset (at least to within the 0.15~magnitude
accuracy of the Abell~2029 zero-point estimation and ignoring the small
Malmquist bias correction).  If offsets are interpreted as due to peculiar
motions, the Abell~2029 offset implies the cluster is essentially at rest
in the CMB frame ($30 \pm 1600$ \kms).  Alternatively, this negligible offset
may be construed as an indication that little evolution in the Tully-Fisher
relation has taken place in clusters up to $z$ \about 0.08.  However, a much
larger sample of clusters at higher redshift is needed to make such a claim
statistically significant.  In any case, this null
result gives us great confidence in the soundness of our data reduction
procedures as it results from the combination of spectroscopic and
photometric data that were obtained and reduced independently for different
projects.
  
\acknowledgements
We would like to thank Riccardo Giovanelli and Jacqueline van Gorkom 
for many helpful comments and
Martha Haynes for her assistance in the preparation of the
observing run.  We thank Steve Boughn for his permission to use the
unpublished $I$~band mosaic observations of Abell~2029.  JMU wishes to
thank Prof. A. Morales for his hospitality and the ``Fundacion BBV'' for its
financial support of a sabbatical leave at the Laboratorio de Fisica Nuclear
of the Universidad de Zaragoza (Spain) during which this paper was completed.
  The results presented here are based on
observations carried out at the Palomar Observatory (PO) and at the Kitt Peak
National Observatory (KPNO).  KPNO is operated by Association of Universities
for Research in Astronomy, Inc., under contract with the National Science
Foundation.  The Hale telescope at the PO is operated by the California
Institute of Technology under a cooperative agreement with Cornell
University and the Jet Propulsion Laboratory.  The NRAO is a facility of the
National Science Foundation which is operated under cooperative agreement by
Associated Universities, Inc.  This research was supported by NSF grant
AST96--17069.

\end{document}